\documentclass[preprint,12pt]{elsarticle}

\usepackage{graphicx}

\usepackage{amssymb}

\journal{Chaos, Solitons \& Fractals}

\begin{document}

\begin{frontmatter}



\title{L\'evy flights in human behavior and cognition}


\author{Andrea Baronchelli}
\address{Laboratory for the Modeling of Biological and Socio-technical Systems, Northeastern University, Boston, MA 02115, USA}

\author{Filippo Radicchi}
\address{Departament d'Enginyeria Quimica, Universitat Rovira i Virgili, Av. Paisos Catalans 26, 43007 Tarragona, Spain}

\begin{abstract}
L\'evy flights represent the best strategy to randomly search for a target in an unknown environment, and have been widely observed in many animal species. Here, we inspect and discuss recent results concerning human behavior and cognition. Different studies have shown that human mobility can be described in terms of L\'evy flights, while fresh evidence indicates that the same pattern accounts for human mental searches in online gambling sites. Thus, L\'evy flights emerge as a unifying concept with broad cross-disciplinary implications. We argue that the ubiquity of such a pattern, both in behavior and cognition, suggests that the brain regions responsible for this behavior are likely to be evolutionarily old (i.e. no frontal cortex is involved), and that fMRI techniques might help to
confirm this hypothesis. 
\end{abstract}


\end{frontmatter}

\section{Introduction}

L\'evy flights describe a class of random walks whose step lengths follow a power-law tailed distribution~\cite{Shlesinger93}. 
In a L\'evy flight, the probability 
$P(d)$ that the walker
performs a jump of length $d$ is given by 

\begin{equation}
P(d) \sim d^{-\gamma},
\label{e:levy}
\end{equation}

\noindent with $1<\gamma \leq 3$. L\'evy flights
have been proven to be the best strategy that can be adopted 
in random searches~\cite{viswanathan1999optimizing, Lomholt2008}. 
In particular, 
Viswanathan and collaborators proved that, independently
of the dimensions of the space where the search is performed,
the optimal exponent is $\gamma=2$ if 
(i) the target sites are sparse, (ii) 
they can be visited multiple times, and (iii) the searcher can detect the target only when it is in its close vicinity \cite{viswanathan1999optimizing}. An important case where these conditions are naturally fulfilled concerns foraging animals looking for scarce preys, and in fact L\'evy flights have been observed in the patterns of movement of different species ranging from  Albatrosses \cite{viswanathan1996levy} to marine predators \cite{sims2008scaling, Humpries2010}, and from monkeys \cite{ramos2004levy} to mussels \cite{de2011levy}. Although in some cases further analysis is needed \cite{edwards2007revisiting}, the overall pattern appears to be robust \cite{viswanathan2008levy}.

The evidence for L\'evy strategies in humans has become manifest in the last few years. Indirect measurement as banknote spreading indicate that  corrected L\'evy flights account for large scale movement of individuals in industrialized countries \cite{brockmann2006scaling}, while the Dobe Ju/'Ohoansi foraging patterns appear to conform more strictly to the L\'evy step distribution \cite{brown2007levy}, as Chilean purse-seiners perform  \cite{bertrand2005levy}. Remarkably, also different cognitive processes have been explained in terms of L\'evy flights. It has been argued that  memory retrieval could be explained in terms of a L\'evy processes \cite{rhodes2007human} , while more recently it was shown that L\'evy flights describe not only movement patterns in real space, but also searches performed by humans in non-physical spaces during online games \cite{radicchi2012rationality}. In addition, this turns out to be the optimal strategy to be adopted, and an evolutionary model accounts for the emergence of this unconscious behavior \cite{radicchi2012evolution}.

The emerging picture is surprising. L\'evy flights underlie many aspects of human dynamics and behavior, and have the potential to become a unifying concept for problems traditionally addressed by different disciplines, such as urban planning, social science and cognitive science. Here we review and discuss the key evidence provided in the recent literature, which is relatively small and scattered across different journals in different disciplines. Our aim is to present a coherent picture of conceptually related findings that may have gone unnoticed to many potentially interested researchers. We also argue that the ubiquity of the L\'evy pattern, both in behavior and cognition of humans and animals, suggests that the brain regions responsible for implementing it are likely to be evolutionarily old (i.e. no frontal cortex is involved), and that fMRI techniques might help shed light on this hypothesis.

\section{L\'evy flights in human mobility and foraging}
\label{}

\subsection{Mobility}
\label{s:macromob}

A neat evidence for the existence of L\'evy flights in human mobility comes from the analysis of online bill-tracking websites, which track the movement of individual banknotes in space and time \cite{brockmann2006scaling}. Motivated by the difficulty to access data about human mobility, the rationale beyond this approach is that banknote displacement mirrors human travel. Brockman et al.  \cite{brockmann2006scaling} analyzed approximately $10^6$ reports of the bill-tracking website {\tt www.wheresgeorge.com} and found that their spatial jumps follows a L\'evy flight  with exponent $\gamma =1.59 \pm 0.02$. Beyond this clear signature for the displacement distribution of human travels, the same analysis highlighted the existence of scale-free periods of rest attenuating the dispersal patterns. Even taking this element into account, however, the observed dynamics remains superdiffusive. In particular the L\'evy flight signature appears to be significantly robust, and independent from the considered observation time windows. While providing a first large scale evidence of a L\'evy behavior in human mobility, this study could not provide all the information required to assess individual movement strategies at small scale. The movement of banknotes, in fact, reflects the combined movement of more individuals who carried the bills between two successive reports to the online tracking system.

More fine-grained insights into human mobility have been gained from the analysis of the mobile phones traces of a set of $10^5$ individuals, whose position were recorded every time their mobile device was actively or passively engaged in a phone call or text message communication \cite{gonzalez2008understanding}. Here, the spatial resolution is determined by the density of the $\sim10^4$ mobile towers, having an average service area of approximately $3$ km$^2$ (but $30\%$ of which with a covered area smaller or equal to $1$ km$^2$). Gonz\'alez and coworkers found a distribution of displacements compatible with (truncated) L\'evy flights, with an exponent of $\beta=1.75 \pm 0.15$, essentially agreeing with the findings of the bank note analysis in  \cite{brockmann2006scaling}. However, the mobile phone data set allows for a deeper investigation of the patterns emerging from the aggregated data. In particular, it turns out that individuals display significant regularities, and spend most of their time in few locations (work, home, etc.). The conclusion is that the truncated L\'evy flight pattern observed in aggregated data, and already revealed by the banknote analysis \cite{brockmann2006scaling},  is the convolution between individual L\'evy trajectories and a population heterogeneity emerging when the radius of gyration of different individuals is considered. 

A further validation of the L\'evy behavior was provided by Rhee and coworkers \cite{rhee2011levy}. The principal merit of their study concerns the unprecedented spatial and temporal resolution at which human mobility has been observed. The analysis relies on the GPS signal of the mobile phones of $101$ individuals, whose position was recorded every $10$ seconds and with an accuracy of $3$ meters in two University Campuses, on theme park, one metropolitan area and one state fair. The total signal amounts to $\sim 2 \times 10^5$ displacement events. The results indicate that both the flight distribution and the pause-time distribution exhibit heavy tail, compatible with a power-law behavior ($1.16 < \gamma < 1.82$, in the different experiments), but also with a Weibull or log-normal distributions. However, a nice corroboration of the L\'evy flight hypothesis comes by a truncated L\'evy flight model that accurately reproduces the experimental curves \cite{rhee2011levy}. It is also interesting noticing that individuals seem to follow a super-diffusive pattern on short time-scales (shorter than one hour) and a sub diffusive one afterwards, even though this is due to individuals reaching the boundary of their individually confined area. Overall, the unparalleled temporal and spatial resolution of this study allows to conclude that L\'evy flights are genuinely intrinsic to human movements, and that, as the author stress, human intentions instead of geographical artifacts play a major role in producing such a behavior \cite{rhee2011levy}.


\subsection{Foraging patterns}


L\'evy flights have been used to explain not only
the movement patterns of individuals, but also
those of groups of individuals. This
has been observed for groups of spider-monkeys \cite{ramos2004levy}
as well as for groups of human  hunter-gatherers \cite{brown2007levy}. 
The movements analyzed by Brown et. al. concern the Dobe band of the Ju/'hoansi population living in deserted areas in Botswana and Namibia\cite{brown2007levy}. This group used to live at the Dobe waterhole during the dry season, while during and after the rains small groups moved out into the hinterland building camps somehow near seasonal water sources where plant food and game was available. When resources were exhausted in the proximity of a camp, they would move again to build another camp or even returning for a short visit to the Dobe waterhole. 
At odds with the studies of section \ref{s:macromob}, the anthropological investigation could count on a much more limited data set, consisting of the locations of the camps that were occupied by a small kin group during approximately $6$ months in $1968$, and documented in \cite{yellen1977archaeological}. In total, this reduced to $37$ displacements and $28$ camps. Notwithstanding this limitation, the analysis of \cite{brown2007levy} shows that the L\'evy flight distribution fits the observed foraging pattern better than a normal, exponential or uniform distributions. Furthermore, the exponents for the length of the displacements is $\gamma = -1.9675$, i.e. extremely close to the optimal value $\gamma=2$ \cite{viswanathan1999optimizing}. Thus, when foraging, humans apparently adopt the same optimal strategy of other animals.

Another evidence in favor of L\'evy flights in human foraging patterns comes from the displacement of Peruvian purse-seiners \cite{bertrand2005levy}. Bertrand et. al. analyzed $68025$ fishing trips of $756$ vessels, equipped with satellite transmitters that record their positions, during $14$ months. The analysis showed that the displacements length follows a clear power law, and aggregating the data for different vessels on a monthly bias provides exponents $ 1.13 < \gamma < 2.09$, with a mode around $\gamma \approx 1.8$. Interestingly the primary goal of the study is to infer the spatial distribution of anchovy, under the assumption that the distribution in space of preys should be somehow mirrored in the foraging pattern of the predator through a a sort of  ``behavioral cascade'' in space \cite{russell1992foraging}. However, the authors do not find any correlation between the observed L\'evy flight exponent and the total surface occupied by the anchovy or with a measure of spatial concentration, but only with a fractal dimension of anchovy distribution that admittedly has ``no intuitive interpretation'' \cite{bertrand2005levy}. In light of the results discussed above, and in particular considering that the L\'evy behavior is optimal in case of a scarce prey  \cite{viswanathan1999optimizing}, we believe that a very likely, as well as simple, explanation is that the L\'evy flight movements of purse-seiners are largely independent of the anchovy distribution, some knowledge of which is more likely to determine small variations at the level of the L\'evy exponent, rather than being responsible for the overall L\'evy behavior.


\section{L\'evy flights in human cognition}

\subsection{Retrieval from semantic memory}

The first study of L\'evy flights in human cognition concerns an experimental investigation of memory retrieval. Nine subjects were asked to recall as many names as possible within a $\sim 20$ minutes span, and their inter-retrieval interval was recorded \cite{rhodes2007human}. Interestingly, Rhodes and Turvey, the authors of the study, observed an exponential increase in the interval between retrievals interrupted by bursts of shorter retrieval intervals. Overall, the retrieval times follow a power law behavior, with exponents $ 1.37 \leq \gamma \leq 1.98$ for the different subjects. The authors propose a suggestive analogy between spatial foraging and memory retrieval, and suggest that ``particular words are randomly and sparsely located in their respective spaces at sites that are not known a priori". The L\'evy behavior would therefore result in an adaptive advantage related to the cognitive power of the \textit{Homo Sapiens} species \cite{viswanathan2008levy}. While this explanation is certainly intriguing, we note that alternative explanations might perhaps account for the observed behavior. In particular, if we consider the process of word recall as a random-walk exploration of a semantic network \cite{baronchelli2013networks}, we see that the task the subjects had to undergo corresponds to the classical problem of network coverage, the experimental focus being on the lag times separating two successive discoveries of new nodes. In this framework, it has been shown that a tree-like topology, for example, is sufficient to trigger the emergence of heavy tailed distributions of lag times \cite{baronchelli2008random}. However, notwithstanding these possible explanations, the original proposal of \cite{rhodes2007human} remains plausible, and further investigations in this context can shed new light on the processes underlying memory coverage. In this respect, the recent theoretical extension of L\'evy flights to complex networks and the evidence that such strategies reduce the coverage time provide new and important conceptual and technical tools \cite{riascos2012long}.

\subsection{Mental searches}

The clearest evidence for a L\'evy flight in nature concerns human mental searches \cite{radicchi2012rationality}. If often, in foraging and mobility studies, ``two orders of magnitude of scaling can represent a luxury'' \cite{viswanathan2008levy}, here the power law decay has in fact been observed over four orders of magnitude. The system under scrutiny consisted of users engaging in Lowest Unique Bids (LUB) on online websites. A LUB is an auction where such prizes as TVs or even cars are won by the participant who offers the lowest and unmatched price. The minimum increment step for bids is one cent. So, for example, if three participants bid on one cent, three of them on two cents, and just one of them on three cents, the latter wins the auction. The revenue for the organizer of the LUB comes from the fees that participants pay for each bid they place. On the other hands, users can win the right to buy the prize for just a few cents. The authors of the present review, together with L.A.N. Amaral, inspected the jumps between consecutive bids in the bidding (one-dimensional) space for $7904$ participants to $580$ auctions in $3$ different websites and found a neat indication that individuals sample a power law distribution with exponent $\gamma \simeq 1.4$, and that \textit{all} the individuals who placed enough bids to be analyzed singularly follow the same behavior. Remarkably, performing a L\'evy flight with that exponent turns out to be the optimal and rational strategy to be adopted in such a competitive arena (i.e., a Nash equilibrium), even though playing a LUB is irrational (no economic gain is expected).

Subsequently, we also showed that a simple evolutionary model is able to account for the emergence of the observed behavior in a population of individuals performing LUBs \cite{radicchi2012evolution}. At the beginning different individuals are endowed with different price-line exploration strategies (i.e. exponents of the sampled jump distribution) and let play the LUB. As in the classical Moran process \cite{nowak2006evolutionary}, the winner of the LUB reproduces transmitting her strategy, possibly slightly mutated, to a new individual (her `offspring') which enters the population replacing a randomly extracted user. Remarkably, it can be shown that the population converges towards a L\'evy flight distribution with the observed exponent irrespective of the initial conditions, and even if at the beginning no individual performs a L\'evy flight (i.e. all exponents are larger than $3$ or smaller than $1$). Thus, the empirically measured L\'evy behavior turns out to be not only a Nash equilibrium, but also an evolutionary stable strategy.

\section{Discussion}

The evidence for L\'evy behavior in human dynamics has affected different disciplines. Understanding individual mobility patterns in modern societies is crucial for the correct characterization and modeling of the spreading of human infectious diseases \cite{brockmann2006scaling}. A correct description of human movements is crucial also to quantify the role of physical proximity on the creation of social networks \cite{gonzalez2008understanding,rhee2011levy}, which in their turn are the substrate for a large spectra of diffusion processes, ranging from mobile routing problems \cite{rhee2011levy}, epidemic or norm spreading \cite{ehrlich2005evolution} to language dynamics \cite{loreto2011statistical}. In the same way, the L\'evy behavior has been largely neglected by archaeologists and anthropologists, but it is likely to have an impact on both theoretical and, through the fractal pattern of movements it implies, field excavations \cite{brown2007levy}.
From the cognitive point of view, on the other hand, identifying the processes underlying memory exploration and retrieval could have a significant impact on the understanding of neurodegenerative diseases \cite{baronchelli2013networks}, while it has been argued that the universality of patterns observed for mental searches could impact not only Cognitive Science, but also Behavioral Economics \cite{ornes2013foraging}.

In conclusion, L\'evy flights appear to be ubiquitous not only among foraging animals, but also in human behavior and cognition, in such different manifestations as individual mobility, group foraging, industrial fisheries and mental searches. It is therefore natural to ask the reason for this ubiquity. Where does the L\'evy conduct come from? It seems clear that the adoption of a L\'evy strategy in both the physical world and in mental representation is beneficial (if not optimal), but it is largely unconscious even for humans. We therefore raised the hypothesis that there is likely no frontal cortex involvement \cite{radicchi2012rationality}. We hope that fMRI techniques will shed light on the nature of the underlying processes responsible for the observed L\'evy behaviors in near future.

\vspace{4cm}

\begin{figure*}[!h]
\begin{center}
\includegraphics[width=0.95\textwidth]{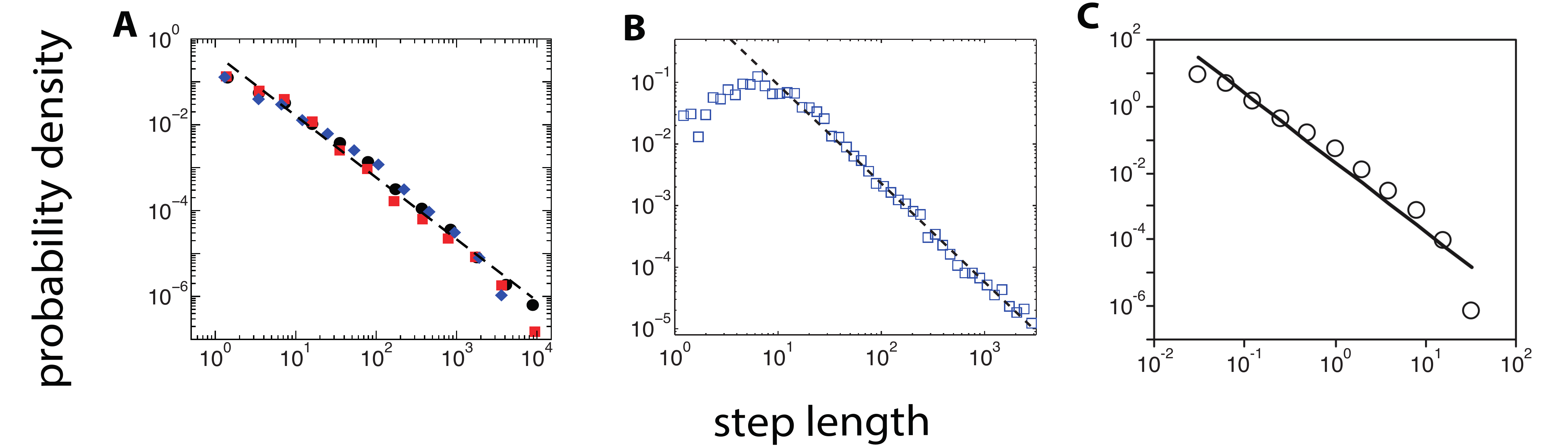}
\end{center}
\caption{Empirical evidence of the
ubiquity of L\'evy flights in natural systems.
{\bf A} Probability density of the difference
between consecutive bid values 
in on-line auctions. Step lengths
are measured in dollars. 
Adapted from Fig.~3A of
Radicchi et al.~\cite{radicchi2012rationality}.
{\bf B} Probability density 
of the distance covered in human travels.
 Step lengths
are measured
in kilometers. 
Adapted from Fig.~1C of
Brockman et al.~\cite{brockmann2006scaling}.
{\bf C} Probability density of the distance between different
positions occupied by Atlantic cod ({\it Gadus morhua}). 
Step lengths
are measured
in meters. Adapted from Fig.~1D of
Sims et al.~\cite{sims2008scaling}.
}
\label{fig}
\end{figure*}

\break






\begin{thebibliography}{10}
\expandafter\ifx\csname url\endcsname\relax
  \def\url#1{\texttt{#1}}\fi
\expandafter\ifx\csname urlprefix\endcsname\relax\def\urlprefix{URL }\fi
\expandafter\ifx\csname href\endcsname\relax
  \def\href#1#2{#2} \def\path#1{#1}\fi

\bibitem{Shlesinger93}
M.~Shlesinger, G.~Zaslavsky, J.~Klafter, Strange kinetics, Nature 363~(6424)
  (1993) 31--37.

\bibitem{viswanathan1999optimizing}
G.~Viswanathan, S.~V. Buldyrev, S.~Havlin, M.~Da~Luz, E.~Raposo, H.~E. Stanley,
  Optimizing the success of random searches, Nature 401~(6756) (1999) 911--914.

\bibitem{Lomholt2008}
M.~Lomholt, K.~Tal, R.~Metzler, K.~Joseph, L{\'e}vy strategies in intermittent
  search processes are advantageous, Proceedings of the National Academy of
  Sciences USA 105~(32) (2008) 11055--11059.

\bibitem{viswanathan1996levy}
G.~Viswanathan, V.~Afanasyev, S.~Buldyrev, E.~Murphy, P.~Prince, H.~E. Stanley,
  L{\'e}vy flight search patterns of wandering albatrosses, Nature 381~(6581)
  (1996) 413--415.

\bibitem{sims2008scaling}
D.~W. Sims, E.~J. Southall, N.~E. Humphries, G.~C. Hays, C.~J. Bradshaw, J.~W.
  Pitchford, A.~James, M.~Z. Ahmed, A.~S. Brierley, M.~A. Hindell, et~al.,
  Scaling laws of marine predator search behaviour, Nature 451~(7182) (2008)
  1098--1102.

\bibitem{Humpries2010}
N.~Humphries, N.~Queiroz, J.~Dyer, N.~Pade, M.~Musyl, et~al., Environmental
  context explains l\'evy and brownian movement patterns of marine predators,
  Nature 465~(7301) (2010) 1066--1069.

\bibitem{ramos2004levy}
G.~Ramos-Fern{\'a}ndez, J.~L. Mateos, O.~Miramontes, G.~Cocho, H.~Larralde,
  B.~Ayala-Orozco, L{\'e}vy walk patterns in the foraging movements of spider
  monkeys (ateles geoffroyi), Behavioral Ecology and Sociobiology 55~(3) (2004)
  223--230.

\bibitem{de2011levy}
M.~de~Jager, F.~J. Weissing, P.~M. Herman, B.~A. Nolet, J.~van~de Koppel,
  L{\'e}vy walks evolve through interaction between movement and environmental
  complexity, Science 332~(6037) (2011) 1551--1553.

\bibitem{edwards2007revisiting}
A.~M. Edwards, R.~A. Phillips, N.~W. Watkins, M.~P. Freeman, E.~J. Murphy,
  V.~Afanasyev, S.~V. Buldyrev, M.~G. da~Luz, E.~P. Raposo, H.~E. Stanley,
  et~al., Revisiting l{\'e}vy flight search patterns of wandering albatrosses,
  bumblebees and deer, Nature 449~(7165) (2007) 1044--1048.

\bibitem{viswanathan2008levy}
G.~Viswanathan, E.~Raposo, M.~Da~Luz, L{\'e}vy flights and superdiffusion in
  the context of biological encounters and random searches, Physics of Life
  Reviews 5~(3) (2008) 133--150.

\bibitem{brockmann2006scaling}
D.~Brockmann, L.~Hufnagel, T.~Geisel, The scaling laws of human travel, Nature
  439~(7075) (2006) 462--465.

\bibitem{brown2007levy}
C.~T. Brown, L.~S. Liebovitch, R.~Glendon, L{\'e}vy flights in dobe ju/Õhoansi
  foraging patterns, Human Ecology 35~(1) (2007) 129--138.

\bibitem{bertrand2005levy}
S.~Bertrand, J.~M. Burgos, F.~Gerlotto, J.~Atiquipa, L{\'e}vy trajectories of
  peruvian purse-seiners as an indicator of the spatial distribution of anchovy
  (engraulis ringens), ICES Journal of Marine Science: Journal du Conseil
  62~(3) (2005) 477--482.

\bibitem{rhodes2007human}
T.~Rhodes, M.~T. Turvey, Human memory retrieval as l{\'e}vy foraging, Physica
  A: Statistical Mechanics and its Applications 385~(1) (2007) 255--260.

\bibitem{radicchi2012rationality}
F.~Radicchi, A.~Baronchelli, L.~A. Amaral, Rationality, irrationality and
  escalating behavior in lowest unique bid auctions, PloS one 7~(1) (2012)
  e29910.

\bibitem{radicchi2012evolution}
F.~Radicchi, A.~Baronchelli, Evolution of optimal l{\'e}vy-flight strategies in
  human mental searches, Physical Review E 85~(6) (2012) 061121.

\bibitem{gonzalez2008understanding}
M.~C. Gonzalez, C.~A. Hidalgo, A.-L. Barabasi, Understanding individual human
  mobility patterns, Nature 453~(7196) (2008) 779--782.

\bibitem{rhee2011levy}
I.~Rhee, M.~Shin, S.~Hong, K.~Lee, S.~J. Kim, S.~Chong, On the levy-walk nature
  of human mobility, IEEE/ACM Transactions on Networking (TON) 19~(3) (2011)
  630--643.

\bibitem{yellen1977archaeological}
J.~E. Yellen, Archaeological approaches to the present: models for
  reconstructing the past, Vol.~1, Academic Press New York, 1977.

\bibitem{russell1992foraging}
R.~W. Russell, G.~L. Hunt~Jr, K.~O. Coyle, R.~T. Cooney, Foraging in a fractal
  environment: spatial patterns in a marine predator-prey system, Landscape
  Ecology 7~(3) (1992) 195--209.

\bibitem{baronchelli2013networks}
A.~Baronchelli, R.~Ferrer-i Cancho, R.~Pastor-Satorras, N.~Chater, M.~H.
  Christiansen, Networks in cognitive science, Trends in Cognitive Science (in
  press). arXiv preprint arXiv:1304.6736.

\bibitem{baronchelli2008random}
A.~Baronchelli, M.~Catanzaro, R.~Pastor-Satorras, Random walks on complex
  trees, Physical Review E 78~(1) (2008) 011114.

\bibitem{riascos2012long}
A.~Riascos, J.~L. Mateos, Long-range navigation on complex networks using
  l{\'e}vy random walks, Physical Review E 86~(5) (2012) 056110.

\bibitem{nowak2006evolutionary}
M.~A. Nowak, Evolutionary dynamics: exploring the equations of life, Harvard
  University Press, 2006.

\bibitem{ehrlich2005evolution}
P.~R. Ehrlich, S.~A. Levin, The evolution of norms, PLoS Biology 3~(6) (2005)
  e194.

\bibitem{loreto2011statistical}
V.~Loreto, A.~Baronchelli, A.~Mukherjee, A.~Puglisi, F.~Tria, Statistical
  physics of language dynamics, Journal of Statistical Mechanics: Theory and
  Experiment 2011~(04) (2011) P04006.

\bibitem{ornes2013foraging}
S.~Ornes, Foraging flights, Proceedings of the National Academy of Sciences
  110~(9) (2013) 3202--3204.

\end{thebibliography}







\end{document}